\documentclass[epsf]{aa}
\usepackage{natbib}
\usepackage{graphicx}

\begin{document}

\title{Superhump-like variation during the anomalous state of SU~UMa}
\subtitle{}
\authorrunning{T. Kato}
\titlerunning{Superhump-like variation during the anomalous state of SU UMa}

\author{Taichi Kato\inst{1}}

\institute{Department of Astronomy, Kyoto University, Kyoto 606-8502, Japan}

\offprints{Taichi Kato, \\ e-mail: tkato@kusastro.kyoto-u.ac.jp}

\date{Received / accepted }

\abstract{
   We observed an anomalously outbursting state of SU~UMa which occurred
in 1992.  Time-resolved photometry revealed the presence of signals
with a period of 0.0832$\pm$0.0019~d, which is 3.6$\sigma$ longer than
the orbital period (0.07635~d) of this system.  We attributed this
signal to superhumps, based on its deviation from the orbital period
and its characteristic profile.  During this anomalous state of SU~UMa,
normal outbursts were almost suppressed, in spite of relatively regular
occurrences of superoutbursts.  We consider that an ensuing tidally
unstable state following the preceding superoutburst can be a viable
mechanism to effectively suppress normal outbursts, resulting in an
anomalously outbursting state.

\keywords{
Accretion, accretion disks  --- novae, cataclysmic variables
           --- Stars: dwarf novae
           --- Stars: individual (SU~UMa)}
}

\maketitle

\section{Introduction}

   SU~UMa is the prototype of SU~UMa-type dwarf novae (for a recent review
of SU~UMa-type stars and their observational properties, see
\citealt{war95suuma}).  Although SU~UMa shows typical superoutbursts and
associated superhumps as in other SU~UMa-type dwarf novae \citep{uda90suuma},
the star is also known to sometimes show anomalous states lacking
superoutbursts, or even normal outbursts (cf. \citealt{ros00suuma}).
According to \citet{ros00suuma},
the period of 1980--1983 was the most remarkable, when SU~UMa almost
completely stopped outbursting.  Several instances have been known,
that SU~UMa showed a lesser degree of anomalously outbursting state.
February, 1992 was another such period (cf. \citealt{ros00suuma}),
when SU~UMa ceased to show normal outbursts.  During that period,
the mean brightness of SU~UMa was observed brighter than the averaged
quiescent level.

\section{Observation}

   We obtained time-resolved CCD photometric runs on three night in 1992
February, when SU~UMa was reported to be in an anomalous state.  The long
observation on February 17 was done between BJD 2448669.959 and 2448670.252,
covering 7 hours.

   Two shorter runs were done under less favorable conditions before and
after this night.  We used a CCD camera (Thomson TH~7882, 576 $\times$ 384
pixels, on-chip 3$\times$3 binning adopted) attached to the Cassegrain
focus of the 60 cm reflector (focal length=4.8 m) at Ouda Station,
Kyoto University \citep{Ouda}.  An interference filter was used
which had been designed to reproduce the Johnson {\it V} band.
The exposure time was 30 s.  The frames were first corrected
for standard de-biasing and flat fielding, and were then processed by a
microcomputer-based aperture photometry package developed by the author.
The magnitudes of the object were determined relative to GSC 4129.95
($V$=13.38, $B-V$=+0.75), whose constancy during the run was confirmed
using GSC 4129.224.  The magnitude of the comparison star is taken from
\citet{mis96sequence}.  Barycentric corrections to the observed times
were applied before the following analysis.  Table \ref{tab:log} lists
the log of observations, together with nightly averaged magnitudes.

\begin{table}
\begin{center}
\caption{Nightly averaged magnitudes of SU~UMa}\label{tab:log}
\begin{tabular}{ccccc}
\hline\hline
Start$^a$ & End$^a$ & Mean mag$^b$ & Error$^c$ & N$^d$ \\
\hline
48868.906 & 48869.190 & $-$0.065 & 0.009 & 119 \\
48869.959 & 48870.252 &    0.315 & 0.006 & 430 \\
48870.925 & 48871.138 &    1.033 & 0.012 & 274 \\
\hline
  \multicolumn{5}{l}{$^a$ BJD$-$2400000.} \\
  \multicolumn{5}{l}{$^b$ Relative magnitude to GSC 4129.9.} \\
  \multicolumn{5}{l}{$^c$ Standard error of nightly average.} \\
  \multicolumn{5}{l}{$^d$ Number of frames.} \\
\end{tabular}
\end{center}
\end{table}

\begin{figure}
  \includegraphics[angle=0,height=6cm]{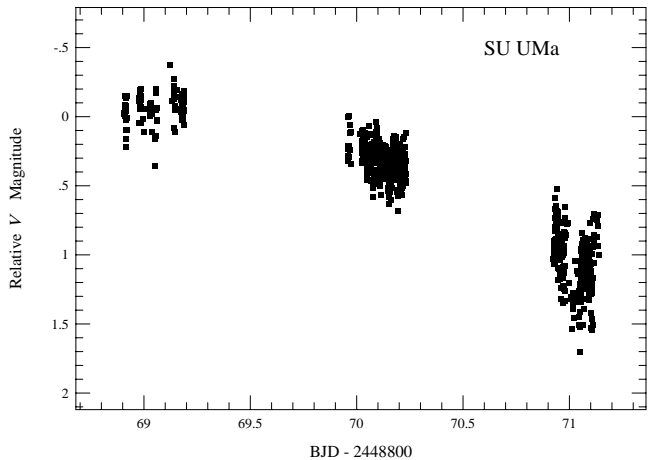}
  \caption{Light curve of a faint outburst of SU~UMa in 1992 February.}
  \label{fig:lc}
\end{figure}

\begin{figure}
  \includegraphics[angle=0,height=6cm]{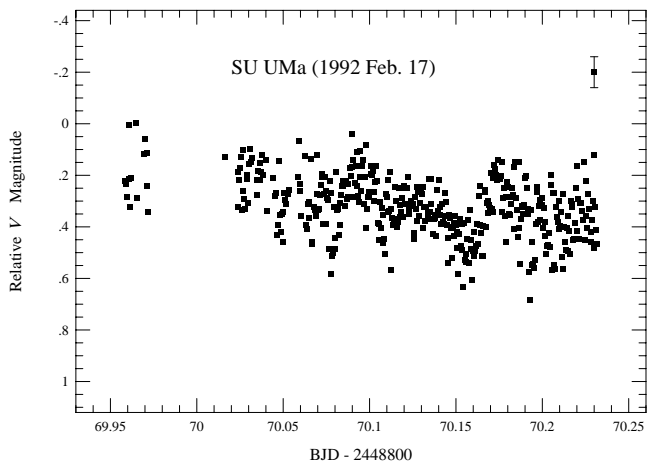}
  \caption{Light curve on 1992 February 17.}
  \label{fig:lc17}
\end{figure}

   Fig. \ref{fig:lc} shows the overall light curve of this observation.
SU~UMa was fading from a small outburst (or a minor brightening),
not apparently recorded in
\citet{ros00suuma}, probably because of the faintness of the peak
brightness ($V\sim$13.3).  Since most of recent outbursts of SU~UMa reach
$V$=12, such a faint outburst was already anomalous.  The enlargement of the
February 17 light curve (Fig. \ref{fig:lc17}) shows hump-like modulations,
with an approximate period close to, but slightly longer than, the orbital
period ($P_{\rm orb}$=0.07635~d, \citet{tho86suuma}).  The observation
on the preceding night was unfortunately affected by a large gap owing to
clouds; on the next night, the object had faded and again moderately
affected by clouds.  Only high-quality February 17 observations were
used for the subsequent period analysis.

\begin{figure}
  \includegraphics[angle=0,height=6cm]{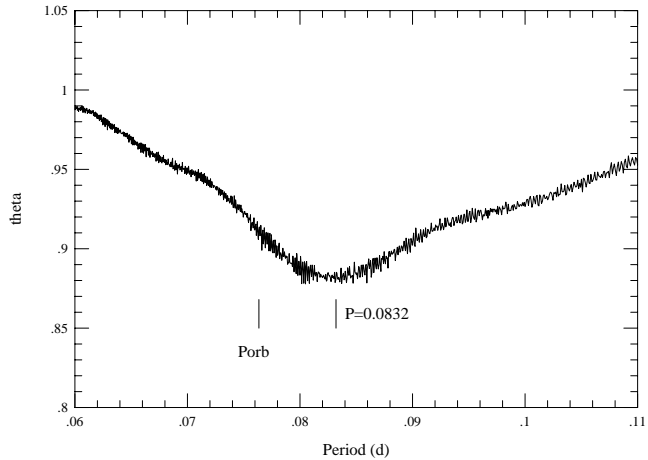}
  \caption{Period analysis of SU~UMa (1992 February 17).}
  \label{fig:pdm}
\end{figure}

   The observations on February 17 were analyzed, after removing a linear
trend of decline, with the Phase Dispersion Minimization (PDM) method
\citep{PDM}.  The resultant theta diagram and phase-averaged
light curve are shown in Figs. \ref{fig:pdm} and \ref{fig:phase},
respectively.

\section{Result and discussion}

   The period analysis strongly supports that presence of the period of
0.0832$\pm$0.0019~d.  The error of the period was estimated using the
application of Lafler-Kinman class of methods by \citet{fer89error}.
Although the error of period estimation is rather
large due to the limited length of a single-night baseline, the period
is 3.6$\sigma$ longer than the orbital period, and is 2.1$\sigma$
longer than the superhump period by \citet{uda90suuma}.  Since superhump
periods can vary to a considerable extent (e.g. \citealt{kat01aqeri}),
the present periodicity, which is significantly longer than the orbital
period, is more regarded as a variety of superhumps, rather than the one
reflecting the orbital period.

\begin{figure}
  \includegraphics[angle=0,height=6cm]{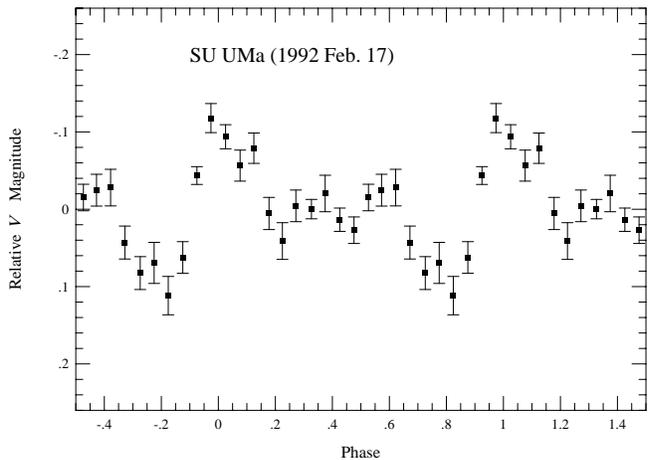}
  \caption{Phase-averaged light curve (1992 February 17).}
  \label{fig:phase}
\end{figure}

The hump profile shown in Fig. \ref{fig:phase}.  A rapid rise and slower
fade, and the presence of a secondary hump at phase 0.6
(cf. \citet{uda90suuma} for the discussion of secondary superhumps
during a superoutburst), are also very characteristic of superhumps.
Such appearance of the superhump signal may be related to the anomalous
state of SU~UMa.

\citet{ros00suuma} argued that the complete cessation of outburst
in the period of 1980--1983 was probably caused by a strong variation
of mass-transfer rates ($\dot{M}$).  This interpretation, however,
does not seem to adequately explain the 1992 anomalous state,
since the interval of superoutbursts (labeled S020 on JD 2448784
and S021 on JD 2449047 in \citet{ros00suuma}) just before and after
this anomalous state was 263 d, which is only slightly longer than the
typical value of this object (see Fig. 5 of \citet{ros00suuma}).
This interval is also a very typical value as seen in other SU~UMa-type
dwarf novae \citep{nog97sxlmi}.
Since the length of a supercycle is primarily governed by the transferred
angular momentum from the secondary star \citep{osa89suuma}, the interval
of successive superoutbursts is roughly inversely proportional to
$\dot{M}$ \citep{ich94cycle}.
This suggests that $\dot{M}$ was relatively normal during this anomalous
state, and that normal outbursts may have been somehow suppressed
even under the condition of a usual $\dot{M}$.

There are some known SU~UMa-type dwarf novae which show a permanently,
or temporarily, reduced frequency of normal outbursts, in contrast
to their high frequency of superoutbursts (V503~Cyg:
\citealt{har95v503cyg}, Ishioka et al. in preparation;
V1113~Cyg: \citealt{kat01v1113cyg}).  The presence of active and inactive
phases in V1113~Cyg, in terms of the frequency of normal outbursts, while
maintaining the supercycle length, seems to require an unknown mechanism
which effectively suppresses normal outbursts \citep{kat01v1113cyg}.

The present discovery of a signal, which can be attributed to superhumps,
during a similarly anomalously outbursting state of SU~UMa provides an
additional clue to this phenomenon.  SU~UMa during anomalous outbursting
states may more or less resemble permanent superhumpers
(cf. \citealt{osa96review}), which do not show strong dwarf nova-type
outbursts, but show superhumps.  The accretion disk in permanent
superhumpers is believed to be hot enough to suppress usual thermal
instability, which is responsible for normal outbursts, but is tidally
unstable, giving rise to permanent superhumps \citep{osa96review}.
Altough it is not yet clear whether such a condition is met during
this anomalously outbursting state of SU~UMa, or whether a similar
explanation can be applicable to SU~UMa-type dwarf novae with strongly
variable activities, the possibility of an ensuing tidally unstable
state following the preceding superoutburst can be a viable mechanism
to effectively suppress normal outbursts.  Since such anomalous states
are known to be rather infrequent, intensive target-of-opportunity
observations are strongly encouraged when future occurrence of such
a state is recognized.

\vskip 3mm

This work is partly supported by a grant-in aid (13640239) from the
Japanese Ministry of Education, Culture, Sports, Science and Technology.

\end{document}